\documentstyle[12pt,epsf,a4,cite]{article}
\begin{document}
%uno due tre quattro
\baselineskip 0.8 true cm

\def\vP{{\mathbf{P}}}
\def\vA{{\mathbf{A}}}
\def\vq{{\mathbf{q}}}
\def\vk{{\mathbf{k}}}
\def\vp{{\mathbf{p}}}
\def\vs{{\mathbf{s}}}
\def\vu{{\mathbf{1}}}
\def\vpu{{\mathbf{p_1}}}
\def\vPu{{\mathbf{P_1}}}
\def\vpd{{\mathbf{p_2}}}
\def\vPd{{\mathbf{P_2}}}
\def\vpt{{\mathbf{p_3}}}
\def\vpq{{\mathbf{p_4}}}
\def\tq{{\mathbf{\widetilde{q}}}}
\def\tpu{{\mathbf{\widetilde{p_1}}}}
\def\tpd{{\mathbf{\widetilde{p_2}}}}
\def\tpt{{\mathbf{\widetilde{p_3}}}}
\def\tpq{{\mathbf{\widetilde{p_4}}}}
\def\vqb{{\mathbf{q_B}}}
\def\vku{{\mathbf{k_1}}}
\def\vkd{{\mathbf{k_2}}}
\def\tqb{{\mathbf{\widetilde{q_B}}}}
\def\tku{{\mathbf{\widetilde{k_1}}}}
\def\tkd{{\mathbf{\widetilde{k_2}}}}
\def\eu{\epsilon_{1}}
\def\ed{\epsilon_{2}}
\def\eub{\epsilon_{1B}}
\def\edb{\epsilon_{2B}}
\def\vkub{{\mathbf{k_{1B}}}}
\def\vkdb{{\mathbf{k_{2B}}}}
\def\vpub{{\mathbf{p_{1B}}}}
\def\vpdb{{\mathbf{p_{2B}}}}
\def\tkub{{\mathbf{\widetilde{k_{1B}}}}}
\def\tkdb{{\mathbf{\widetilde{k_{2B}}}}}

\begin{center}
{ \Large \bf
Polarization effects in elastic electron-proton scattering \\
(Lecture notes)}

{\rm\large  Michail P. Rekalo \footnote{ Permanent address:
\it National Science Center KFTI, 310108 Kharkov, Ukraine}
and Egle Tomasi-Gustafsson}

{\it DAPNIA/SPhN, CEA/Saclay, 91191 Gif-sur-Yvette Cedex,
France}

\today

\end{center}

%\maketitle
\begin{abstract}
In these notes we present a detailed and complete scheme for the calculation of polarization phenomena in elastic electron-nucleon scattering. The longitudinal and transversal components of the polarization for the scattered nucleon, induced by the scattering of longitudinally polarized electrons and the asymmetries for the collision of polarized electrons with a polarized nucleon target are included. The calculations are done in the framework of the one-photon mechanism, in the Breit system, which is the most convenient for the analysis of polarization effects. We briefly discuss the region of applicability of such model and other mechanisms (radiative and electroweak corrections) which
may come into this idealized picture. 
\end{abstract}

\section{Introduction}

Polarization phenomena in elastic $eN$-scattering, $e+N\to e+N$, $N=p$ or $n$, are very peculiar, due to the applicability of the one-photon mechanism and to the fact that nucleon electromagnetic form factors (FF) are real functions of the momentum transfer square,  in the space-like region. As a result, all one-spin polarization observables (being T-odd and P-even) are identically zero at any electron energy and scattering angle. The simplest non-zero polarization observables are two-spin correlations, being P- and T-even. Generally, these observables are characterized by large absolute values and a weak dependence on the electron energy and on the  momentum transfer squared, whereas the differential cross section (with unpolarized particles) shows a very steep decrease with the momentum transfer. 

The calculation of all possible polarization observables for elastic $eN$-scattering can be done exactly (in terms of nucleon electromagnetic FFs) in a model independent way, using the well established formalism of QED.

Polarization phenomena for elastic $eN$-scattering bring essential information in different physical problems. Let us mention some of them.
\begin{itemize}
\item The scattering of longitudinally polarized electrons by a polarized nucleon target ($\vec e+\vec N\to e+N$) or the measurement of the polarization of the final nucleon in $\vec e+ N\to e+\vec N$ can be considered as a very effective method 
\cite{Re68,Do69,Re74,Ar80} for the precise measurement of the electric nucleon FF, $G_{EN}$, at any momentum transfer - in case of neutron, and for large momentum transfer - in case of proton. This method was successfully realized at Bates \cite{Mi98} and JLab \cite{Jo00}.
\item The scattering of relativistic polarized protons by polarized atomic electrons is a possible method to measure the polarization of high energy proton beams, which is essentially energy independent \cite{Ni98}.

\item The scattering of stored proton beams, accumulated in storage rings, by polarized internal hydrogen gas target induces polarization of protons, (initially unpolarized) after millions of turns \cite{Ra93}. The contribution of $p+\vec e\to \vec p+ e$-scattering has also to be taken into account \cite{Ho94}. This method of polarizing hadronic beams seems to offer a realistic scheme to produce polarized antiproton beams. Note in this connection, that polarization phenomena for $p+e$- and $\overline{p}+e$-elastic scattering are the same, in the one-photon approximation.
\end{itemize}

These different applications indicate that the polarization phenomena for elastic $ep$-scattering have to be considered in different frames:

\begin{itemize}
\item Laboratory system (Lab): for elastic scattering of a relativistic electron beam by a nucleon target (nucleon in rest);
\item Laboratory system  in inverse kinematics: for the scattering of relativistic proton beams by atomic electrons;
\item Colliding asymmetrical regimes: for collisions of relativistic beams of electrons and protons (HERA), where each beam has a different energy.
\end{itemize}

In principle two schemes can be suggested to unify the analysis of polarization phenomena in different coordinate systems. 

One is based on the relativistic description \cite{Bj79} of the polarization properties of the Dirac particles: it is possible to find exact formulas for all polarization effects in general relativistic invariant form \cite{Ak58}, as combinations of products of 
four-vector of polarizations and momenta of particles. These general formulas allow to analyze different kinematical situations, substituting the concrete expression for the momenta and polarization four-vectors.

A second way, more simple and elegant, is the calculation of polarization observables in a specific coordinate system, the Breit system, where the analytical formulas for polarization observables have a transparent physical meaning. Expressions are simplified and their validity is easier to be tested, at different steps.
This system effectively exploits the properties of the one-photon mechanism.  Therefore, for elastic $ep$-scattering, the Breit system can be considered as the analogue of the center of mass system (CMS) for the annihilation reaction  $e^++e^-\to p+\overline{p}$.

In particular the definition of the electric $G_{EN}$ and magnetic $G_{MN}$ nucleon FFs enters naturally in the Breit system for the nucleon electromagnetic current. The space representation of the nucleon structure, as Fourier transform of the Sachs FFs, $G_{EN}$ and $G_{MN}$ is valid only in the Breit system, at any value of momentum transfer.

As a rule, the effect of a transversal electron polarization is very small, being decreased by the factor $\gamma^{-1}=m_e/\epsilon$, where $\epsilon$ is the electron energy. But any polarization state of the target plays an  equivalent role. This holds also for the scattered nucleon, which has longitudinal and transversal polarizations, for values of  momentum transfer
$(-q^2)\le$ 10 GeV$^2$. 

But it is not the case for the scattering of relativistic protons on atomic electrons: here all polarization directions of the target electron are important, but, for proton, the longitudinal polarization plays the main role.

These notes focus on the calculation of cross section and polarization phenomena for elastic $eN$-scattering, with application to the scattering of longitudinally polarized relativistic electrons beams on a nucleon target in rest. The possibility to neglect the electron mass simplifies essentially the calculations. We present a scheme of the exact calculation of two-spin polarization phenomena for elastic $eN$-scattering, for two types of experiments:
$\vec e+\vec N\to e+N$ and $\vec e+N\to e+\vec N$. All calculations are done in the framework of the one-photon mechanism, without including effects of radiative corrections or electroweak corrections (due to $Z^0$-boson exchange).

\section{Kinematics in the Breit system}

\begin{figure}[h]
\hspace*{3true cm}
\mbox{\epsfxsize=8.cm\leavevmode\epsffile{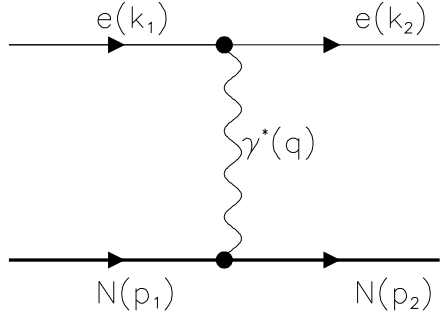}}
%\vspace*{-2truecm}
\caption{One-photon exchange diagram for elastic scattering, $e+N\to e+N$.}
\label{fig:epel}
\end{figure}

The Feynman diagram for elastic $eN$-scattering is shown in Fig. \ref{fig:epel}, assuming one-photon exchange. The notations of the particle four-momenta are also shown in the Fig. 1 and in Table 1 (we will use, in our calculation, the system where $\hbar$=c=1).

%\vspace*{-2truecm}

The conservation of four-momenta at each vertex of the considered diagram can be written as:
\begin{equation}
q=k_1-k_2=p_2-p_1,
\label{eq:eq1}
\end{equation}
and it is valid in any reference frame. Using the relation (\ref{eq:eq1}) in the Lab-system, we derive the formula for the momentum transfer squared $q^2$, which is the basic kinematical variable for elastic $eN$ scattering:
$$q^2=(p_2-p_1)^2=p_1^2+p_2^2-2mE_2 =2m^2-2mE_2=-2mT,$$ 
where $E_2(T)$ is the total (kinetic) energy of the final nucleon, $m$ is the nucleon mass, and $T=E_2-m$.
This formula demonstrates that, for elastic scattering, the momentum transfer squared, $q^2$, is negative for any energy and scattering angle of the outgoing electron. This is true in any reference system, $q^2$ being a relativistic invariant. The kinematical region for which $q^2<0$ is called the $space-like$ region.

\begin{center}

\begin{tabular}{|c|c|c|r|}
\hline
%&&&\\
&\textbf{Lab}&\textbf{CMS}&\textbf{Breit}\\
%&&&\\
\hline
$q$ & $(\omega,\vq )$& $(\widetilde{\omega},\tq )$& $(\omega_B=0,\vqb)$\\
$k_1$ & $(\epsilon_1,\vku)$& $(\widetilde{\epsilon_1},\tku)$& $(\epsilon_{1B},\vkub)$\\
$p_1$ & $(m,0 )$& $(\widetilde{E_1},-\tku )$& $(E_{1B},\vpub)$\\
$k_2$ & $(\epsilon_2,\vkd)$& $(\widetilde{\epsilon_2},\tkd)$& $(\epsilon_{2B},\vkdb)$\\
$p_2$ & $(E_2,\vpd)$& $(\widetilde{E_2},-\tkd)$& $(E_{2B},-\vpub)$\\
\hline
\end{tabular}

\vspace*{.2 true cm}
{\small {\bf Table 1.} Notation of four-momenta in different reference frames.}
\end{center}

\subsection{Proton kinematics in the Breit system}

The most convenient frame for the analysis of elastic $eN$-scattering is the Breit frame, which is defined as the system where the initial and final nucleon energies are the same. As a consequence, the energy of the virtual photon vanishes and its four-momentum squared, $q^2$, coincides with its three-momentum squared,  $\vqb^2$, more exactly, $q^2=-\vqb^2$.
The derivation of the formalism in Breit system is therefore more simple and 
has some analogy with a non-relativistic description of the nucleon electromagnetic structure. From the energy conservation, and from the definition of the Breit system, one can find:
$$\omega_B=E_{1B}-E_{2B}=0,$$
where all kinematical quantities in the Breit system are denoted with subscript $B$. The proton three-momentum can be found from the relation 
$$E_{1B}^2=E_{2B}^2=\vpub^2+m^2=\vpdb^2+m^2,~i.e.~\vpub^2=\vpdb^2.$$ 
The physical solution of this quadratic relation is 
$\vpub=-\vpdb$, as the Breit system moves in the direction of the outgoing proton.
From the three-momentum conservation, in the Breit system $\vqb+\vpub=\vpdb$, one can find: $$\vpub=-\displaystyle\frac{\vqb}{2},
~~\vpdb=\displaystyle\frac{\vqb}{2}.$$
The proton energy can be expressed as a function of $\vqb^2$, and therefore of $q^2$:
$$E_{1B}^2=E_{2B}^2=m^2+
\displaystyle\frac{\vqb^2}{4}=m^2-
\displaystyle\frac{q^2}{4}=m^2(1+\tau),$$
where we replaced the three-momentum in Breit system by the four-momentum  and we introduced the dimensionless quantity $\tau=-\displaystyle\frac{q^2}{4m^2}\ge 0$.

\subsection{Electron kinematics in the Breit system}

\begin{figure}[h]
\hspace*{2true cm}\mbox{\epsfxsize=10.cm\leavevmode \epsffile{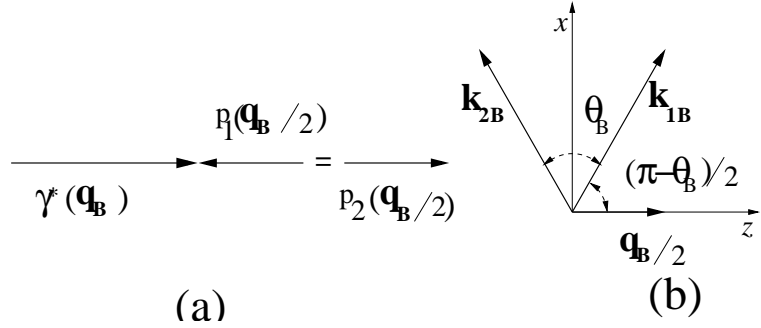}}
\caption{Proton (a) and electron (b) three-momenta representation for elastic $eN$-scattering in the Breit system.}
\label{fig:Breit}
\end{figure}

The conservation of the four momentum, at the electron vertex, can be written, in any reference system, as:
$k_1=q+k_2$ (the virtual photon is radiated by the electron).
In the Breit system, the energy and momentum conservation is:
\begin{equation}
\hspace*{3truecm}
\left\{ \begin{array}{ll}
         \eub=&\omega_B+\edb = \edb, \\
        \vkub=&\vqb+ \vkdb.          
\end{array}
\right .
\end{equation}	  
In order to proceed further, we must define a reference (coordinate) system: we choose the $z$-axis parallel to the photon three-momentum in the Breit system: $z\parallel\vqb$, and the $xz$-plane as the scattering plane.
So we can write:
\begin{equation}
\left\{ \begin{array}{ll}
         \eub^2&=\edb^2=m_e^2+(k_{1B}^{x})^2+(k_{1B}^z)^2=
	 m_e^2+(k_{2B}^x)^2+(k_{2B}^z)^2\\
         k_{1B}^{x}=&k_{2B}^x \\
         k_{1B}^y=&k_{2B}^y=0 \\
	      k_{1B}^z=& q_B+k_{2B}^z \\         
\end{array}
\right .
\end{equation}
It follows $k_{1B}^z=-k_{2B}^z=\displaystyle\frac{q_B}{2}$ (the other possible solution $k_{1B}^z=k_{2B}^z$ would imply $q_B=0$). A graphical representation for the conservation of three-momenta is given in Fig. \ref{fig:Breit}.

We can then write, for the components of the initial and final electron three-momenta:
\begin{equation}
\vkub=(k_{1B}^x,k_{1B}^y,k_{1B}^z)=\left( \displaystyle\frac{q_B}{2}\cot\displaystyle\frac{\theta_B}{2},0, \displaystyle\frac{q_B}{2}\right )= \displaystyle\frac{\sqrt{-q^2}}{2}
\left(\cot\displaystyle\frac{\theta_B}{2},0,1\right ),
\label{eq:vkube}
\end{equation}
\begin{equation}
\vkdb=(k_{2B}^x,k_{2B}^y,k_{2B}^z)=\left( \displaystyle\frac{q_B}{2}\cot\displaystyle\frac{\theta_B}{2},0, -\displaystyle\frac{q_B}{2}\right )= \displaystyle\frac{\sqrt{-q^2}}{2}
\left(\cot\displaystyle\frac{\theta_B}{2},0,-1\right ),
\label{eq:vkdbe}
\end{equation}
The energy of the electron is (in the limit of zero electron mass) is given by:
$$\eub^2=\vkub^2=(k_{1B}^x)^2+(k_{1B}^z)^2=\displaystyle\frac{-q^2}{4\sin^2
\displaystyle\frac{\theta_B}{2}}~\mbox{and}~ \edb=\eub.$$

\subsection{Relation between the electron scattering angles in the Lab system, $\theta_e$ and in the Breit system, $\theta_B$}

As the Breit system is moving along the $z$-axis,  the $x$ and $y$ components of the particle three-momenta do not change after transformation from the Lab to the Breit system: 
\begin{equation}
\left\{ \begin{array}{ll}
   k_{1y}^B=&k_{2y}=0\\
   k_{1x}^B=&k_{1x}.
\end{array}
\right .
\label{eq:eqa}
\end{equation}
From $\vku ^2=k_{1x}^2+k_{1z}^2$ one can find:
\begin{equation}
k_{1x}^2=\vku ^2- \displaystyle\frac{(\vku\cdot\vq)^2}{\vq ^2}
=\displaystyle\frac{\vku ^2\vq ^2- (\vku\cdot\vq)^2}{\vq ^2}
=\displaystyle\frac{\eu^2\ed^2\sin^2\theta_e}{\vq ^2}=
\displaystyle\frac{4\eu^2\ed^2}{\vq ^2}\sin^2
\displaystyle\frac{\theta_e}{2}\cos^2
\displaystyle\frac{\theta_e}{2},
\label{eq:eq7}
\end{equation}
where we replaced $\vq=\vku-\vkd$, $\vku^2=\eu^2$, $\vkd^2=\ed^2$ after setting $m_e=0$.
On the other hand we find for the square of the four-vector $q^2$, the following expression in the Lab system (in terms of the energies of the initial and final electron and of the electron scattering angle):
$$
q^2=(k_1-k_2)^2=
2m_e^2-2k_1\cdot k_2\stackrel{m_e=0}{\simeq}  -2 \eu\ed+2\vku\cdot\vkd=-2\eu\ed(1-\cos\theta_e)= $$
\begin{equation}
-4\eu\ed\sin^2
\displaystyle\frac{\theta_e}{2}.
\label{eq:eq8}
\end{equation}
Comparing Eqs. (\ref{eq:eq7}) and (\ref{eq:eq8}), we find:
$$k_{1x}^2=\displaystyle\frac{(q^2)^2}{4\vq ^2}\cot^2\displaystyle\frac{\theta_e}{2}.$$
Using the relations: $\vq ^2=\omega^2-q^2$ and $q^2+2q\cdot p_1+p_1^2=p_2^2$, we have, in the Lab system, $\omega=-\displaystyle\frac{q^2}{2m}$ and $\vq ^2=-q^2(1+\tau)$. Finally: $$k_{1x}^2=-\displaystyle\frac{q^2}{4(1+\tau)}\cot^2\displaystyle\frac{\theta_e}{2}$$
So, from the relation $k_{1x}^2=(k_{1B}^x)^2$, we find the following relation between the electron scattering angle in the Lab system and in the Breit system:
\begin{center}
\fbox{$
\cot^2\displaystyle\frac{\theta_B}{2}=\displaystyle\frac{\cot^2\theta_e/2}{1+\tau}.
$
}
\end{center}
\vspace*{-1true cm}
\begin{equation}
\label{eq:cot}
\end{equation}

\section{Dynamics}

The elastic $eN$ scattering involves four particles, with spin 1/2. The relativistic  description of the spin properties of each of these particles is based on the Dirac equation:
$$(\hat k-m)u(k)=0,~~\hat k=E\gamma_0-\vk \cdot\vec\gamma,$$
where $k$ is the particle four momentum ($k=(E,\vk)$) and $u(k)$ is a four-component Dirac spinor. We shall use the following representation of the Dirac $4\times 4$ matrixes:
\begin{equation}
\gamma_0=\left (
\begin{array}{ll}
   1&~0\\
   0 &-1
\end{array}
\right ),~
\vec\gamma=\left (
\begin{array}{ll}
   ~0&\vec\sigma\\
   -\vec\sigma &0 
\end{array}
\right ),
\label{eq:g0}
\end{equation}
where $\vec\sigma$ is the standard set of the Pauli $2\times 2$ matrixes. On the basis of the Dirac equation one can write:
\begin{equation}
u(k)=\sqrt{E+m}
\left (\begin{array}{l}
             \chi\\
            \displaystyle\frac{\vec\sigma\cdot\vk }{E+m}\chi
\end{array}
\right ),
\label{eq:gk}
\end{equation}
where $\chi$ is a two-component spinor. We used here the relativistic invariant normalization for the four-component spinors, $u^\dagger u=2E$. 

Two vertexes are present in Fig. \ref{fig:epel}: (1) the electron vertex, which is described by QED-rules, (2) the proton vertex described by QCD and hadron electrodynamics.
The matrix element corresponding to this diagram, is written as:
\begin{equation}
{\cal M}=\displaystyle\frac{e^2}{q^2} 
\ell_{\mu}{\cal J}_{\mu}=\displaystyle\frac{e^2}{q^2} 
\ell\cdot {\cal J},
\label{eq:mat}
\end{equation}
where $\ell_{\mu}=\overline{u}(k_2)\gamma_{\mu}u(k_1)$ is the electromagnetic current of electron. The nucleon electromagnetic current, ${\cal J}_{\mu}$ describes the proton vertex and is generally written in terms of Pauli and Dirac FFs $F_1$ and $F_2$:
\begin{equation}
{\cal J}_{\mu}=\overline{u}(p_2)\left [F_1(q^2)\gamma_{\mu}-
\displaystyle\frac{\sigma_{\mu\nu}q_\nu}{2m}F_2(q^2)\right ]u(p_1),
\label{eq:eqj}
\end{equation}
with
$$\sigma_{\mu\nu}=\displaystyle\frac{\gamma_{\mu}\gamma_{\nu}-
\gamma_{\nu}\gamma_{\mu}}{2}.$$
Note that ${\cal J}\cdot q=0$, for any values of $F_1$ and $F_2$, i.e. the current ${\cal J}_{\mu}$ is conserved\footnote{This can be easily proved as follows. The term $\sigma_{\mu\nu}q_{\mu}q_{\nu}$ vanishes, because it is the product of a symmetrical and antisymmetrical  tensors, and $\overline{u}(p_2)\hat q u(p_1)=\overline{u}(p_2)(\hat{p}_2-\hat p_1)u(p_1)
=\overline{u}(p_2)(m-m)u(p_1)=0$, as a result of the Dirac equation for both four-component spinors, $u(p_1)$ and $u(p_2)$. Note that the current (\ref{eq:eqj}) is conserved only when both  nucleons ( in initial and final states) are real, the form factor $F_1$ violates the current conservation, if one nucleon is virtual}.

Using the Dirac equation for the four-component spinors of the initial and final nucleon, Eq. (\ref{eq:eqj}) can be rewritten in a simpler form (see Appendix I):
\begin{equation}
{\cal J}_{\mu}=\overline{u}(p_2)\left [ \left ( F_1+F_2\right ) \gamma_{\mu}
- \displaystyle\frac{(p_1+p_2)_\mu}{2m}F_2\right ] u(p_1),
\label{eq:jp}
\end{equation}
which is also conserved. 

Eq. (\ref{eq:jp}) (as well as (\ref{eq:eqj})) is the expression of the nucleon electromagnetic current, which holds in any reference system. However, for the analysis of polarization phenomena, the Breit system is the most preferable. First of all, the explicit expression of the current 
${\cal J}_{\mu}=({\cal J}_0,\vec{\cal J})$ is simplified in the Breit system:
\begin{equation}
\left\{ \begin{array}{ll}
         {\cal J}_0&= \overline{u}(p_2)\left [ \left ( F_1+F_2\right ) \gamma_0
- \displaystyle\frac{(E_{1B}+E_{2B})}{2m}F_2\right ] u(p_1),~E_{1B}=E_{2B}=E, \\
\vec{\cal J}=&\overline{u}(p_2)\left [ \left ( F_1+F_2\right ) \vec\gamma
  - \displaystyle\frac{(\vpub+\vpdb)}{2m}F_2\right ] u(p_1)=
 \left ( F_1+F_2\right ) \overline{u}(p_2) \vec\gamma u(p_1).\\
\end{array}
\right .
\end{equation}
With $u(p_1)$ and $u(p_2)$ defined according to (\ref{eq:gk}) we find, for the time component ${\cal J}_0$ of the current ${\cal J}_{\mu}$:
\begin{equation}
 \begin{array}{ll}
         {\cal J}_0&= \left (F_1+F_2\right )u^\dagger(p_2)u(p_1)-F_2
	 \displaystyle\frac{E}{m} u^\dagger(p_2)\gamma_0u(p_1)\\
	 &=(E+m)\left \{ \left ( F_1+F_2\right )\chi_2^\dagger\left 
	 (1~,~ \displaystyle\frac{\vec\sigma\cdot\vqb}{2(E+m)}\right )
\left (\begin{array}{l}
             \chi_1\\
            \displaystyle\frac{-\vec\sigma\cdot\vqb}{2(E+m)}\chi_1
\end{array}
\right ) \right .\\
&-F_2	\displaystyle\frac{E}{m}\chi_2^\dagger\left ( 1~,~  \displaystyle\frac{\vec\sigma\cdot\vqb}{2(E+m)}\right )
\left [\begin{array}{ll}
             1&0\\
             0&-1
\end{array}
\right ]
\left .\left (\begin{array}{l}
             \chi_1\\
            \displaystyle\frac{-\vec\sigma\cdot\vqb}{2(E+m)}\chi_1
\end{array}
\right )\right \}=\\
&= 2m\chi_2^\dagger\chi_1\left (F_1-\tau F_2\right ),
\end{array}
\label{eq:eqp1}
\end{equation}
where we used the definition:
$$\vpdb^2=E^2-m^2=\displaystyle\frac{\vqb^2}{4},\mbox{ so that } \displaystyle\frac{\vqb ^2}{4(E+m)^2}=\displaystyle\frac{E-m}{E+m},$$
and
$$\overline{u}(p_2)=u^\dagger(p_2)\gamma_0,~\gamma_0^2=1 \mbox{~and~}  (\vec\sigma\cdot\vq )(\vec\sigma\cdot\vq )=\vq^2.$$

For the vector part $\vec{\cal J}$ of the nucleon electromagnetic current we can find similarly:
\begin{equation}
\begin{array}{ll}
         \vec {\cal J}&= \left (F_1+F_2\right )(E+m)
	\chi_2^\dagger\left ( 1~,~-\displaystyle\frac{\vec\sigma\cdot\vqb }{2(E+m)}\right ) 
\left [ \begin{array}{ll}
             0&\vec\sigma\\
             -\vec\sigma &0\\
\end{array}
\right ]
\left (\begin{array}{l}
             \chi_1\\
            \displaystyle\frac{-\vec\sigma\cdot\vqb }{2(E+m)}\chi_1\\
\end{array}
\right )=\\
&-\displaystyle\frac{1}{2} \left (F_1+F_2\right )\chi_2^\dagger
\left (\vec\sigma\vec\sigma\cdot\vqb -\vec\sigma\cdot\vqb \vec\sigma\right )=\left (F_1+F_2\right )i\chi_2^\dagger\vec\sigma\times\vqb \chi_1.
\end{array}
\label{eq:eqp2}
\end{equation}
Finally:
\begin{center}
\fbox{$
\begin{array}{ll}
{\cal J}_0&= 
 2m\chi_2^\dagger\chi_1\left (F_1-\tau F_2\right )\\
\vec {\cal J}&=
i\chi_2^\dagger\vec\sigma\times\vqb \chi_1\left (F_1+F_2\right )
\end{array}
$
}
\end{center}

These expressions for the different components of the current ${\cal J}_{\mu}$ are valid in the Breit frame only, and allow to introduce in a straightforward way the Sachs nucleon electromagnetic FFs \cite{Sachs}, electric and magnetic, which are written as:
$$G_{EN}=F_1-\tau F_2 ~~G_{MN}=F_1+F_2.$$
Such identification can be easily understood, if one takes into account that the time component of the current, ${\cal J}_0$, describes the interaction of the nucleon electric charge with the Coulomb potential. Correspondingly, the space component 
$\vec {\cal J}$ describes the interaction of the nucleon spin with the magnetic field. 

\section{Unpolarized cross section: the Rosenbluth formula}

The unpolarized cross section is calculated in the CMS system as:
$$
\left (\displaystyle\frac{d\sigma}{d\Omega}\right )_{CMS}=\displaystyle\frac
{\overline {\left | {\cal M}\right |}}{64 \pi^2s}
\displaystyle\frac{q_i}{q_f},
$$
where $s=(k_1+p_1)^2$ is the square of the total energy for the colliding particles. The line denotes the summing over the polarizations of the final particles and the averaging over the polarizations of the initial particles.

From Eq.(\ref{eq:mat}) we can find the following representation for $\overline {\left | {\cal M}\right |}$
\begin{equation}
\overline{\left |{\cal M}\right |^2}=\left (\displaystyle\frac{e^2}{q^2}
\right )^2\overline{\left |\ell\cdot {\cal J}\right|^2}=\left (\displaystyle\frac{e^2}{q^2}
\right )^2L_{\mu\nu}W_{\mu\nu},
\label{eq:cscm}
\end{equation}
where:

$L_{\mu\nu}=\overline{\ell_{\mu}\ell_{\nu}^*}$ is the leptonic tensor;

$W_{\mu\nu}=\overline{{\cal J}_{\mu}{\cal J}_{\nu}^*}$ is the hadronic tensor.

The product of the tensors $L_{\mu\nu}$ and $W_{\mu\nu}$ is a relativistic invariant, therefore it can be calculated in any reference system.  The differential cross section, in any coordinate system, is can be expressed in terms of the matrix element as:
\begin{equation}
d\sigma=\displaystyle\frac{(2\pi)^4\overline{\left|{\cal M}\right |^2}}{4\sqrt{(k_1\cdot p_1)^2-m_e^2m^2}}\delta^4(k_1+p_1-k_2-p_2)
\displaystyle\frac{d^3\vkd}{(2\pi)^3 2\epsilon_2}
\displaystyle\frac{d^3\vpd}{(2\pi)^3 2E_2}.
\label{eq:csma}
\end{equation}
To compare with experiments, it is more convenient to use the differential cross section in Lab system,
${d\sigma}/{d\Omega}_e$, where $d\Omega_e$ is the element of the electron solid angle in the Lab system.
This can be done, integrating Eq. (\ref{eq:csma}), using the properties of the $\delta^4$ function.

First of all, let us integrate over the three-momentum $\vpd$, applying the three momentum conservation for the considered process:
$$ \int d^3\vpd \delta^3(\vku-\vkd-\vpd)=1,\mbox{~with ~the ~condition~} \vpd=\vku-\vkd.$$
Using the definition $d^3\vkd\stackrel{m_e=0}{=}d\Omega_e\vkd^2d|\vkd|\simeq d\Omega_e\ed^2d\ed$, we can integrate over the electron energy, taking into account the conservation of energy:
$$\delta\left (\eu+m-\ed-E_2)d\ed=\delta(\eu+m-\ed-\sqrt{m^2+\vpd^2} \right )d\ed=$$
$$\delta\left (\eu+m-\ed-\sqrt{m^2+(\vku-\vkd)^2}\right )d\ed $$
Let us recall that:
$$\int \delta\left 
[ f(\ed)\right ]d\ed = 
\displaystyle\frac{1}{|f^\prime (\ed)|},$$
where $f(\ed)=\eu+m-\ed-\sqrt{m^2+\eu^2+\ed^2-2\eu\ed\cos\theta_e}$.
Therefore:
$$|f^\prime(\ed )|=1+\displaystyle\frac{\ed-\eu\cos\theta_e}{E_2}=
1+\displaystyle\frac{\ed^2-\vku\cdot\vkd}{\ed E_2}=
\displaystyle\frac{k_2\cdot (k_1+p_1)}{\ed E_2},$$
where we multiplied by $\ed $ the numerator and denominator, and we used the conservation of energy $\ed +E_2=\eu+m$. But from the conservation of four-momentum, in the following form $k_1+p_1-k_2=p_2$, we have:
$$(k_1+p_1)^2+k_2^2-2(k_1+p_1)\cdot k_2=m^2.$$
So $2(k_1+p_1)\cdot k_2=(k_1+p_1)^2-m^2=2k_1\cdot p_1=2\eu m$ (in Lab system).
Finally  
$$|f^\prime(\ed)|= \displaystyle\frac{\eu}{\ed}\displaystyle\frac{m}{E_2}.$$
After substituting in Eq. (\ref{eq:csma}), one finds the following relation between $\overline{\left| {\cal M}\right |^2}$ and the differential cross section in Lab system:
\begin{equation}
\displaystyle\frac{d\sigma}{d\Omega}_e=\displaystyle\frac
{\overline{\left| {\cal M}\right |^2}}{64\pi^2}
\left (\displaystyle\frac{\ed}{\eu}\right )^2
\displaystyle\frac{1}{m^2}.
\label{eq:csm2}
\end{equation}

\subsection{Hadronic tensor $W_{\mu\nu}$}

Let us calculate the hadronic tensor $W_{\mu\nu}$ in the Breit system, where there is a simple expression of the nucleon current. Let us write this current as:
${\cal J}_{\mu}=\chi_2^\dagger F_{\mu}\chi_1$,
with $F_{\mu}=2mG_{EN}$, for $\mu=0$ and  $F_{\mu}=i\vec\sigma\times\vqb G_{MN}$, for $\mu=x,y,z$. So the  the four components of $F_{\mu}$, in terms of the FFs
$G_{EN}$ and $G_{EN}$, can be written as:
\begin{equation}
F_{\mu}=\left\{ \begin{array}{ll}
         2mG_{EN}&, \mu=0 \\
         i\sqrt{-q^2} G_{MN}\sigma_y&, \mu=x \\
     -i\sqrt{-q^2} G_{MN}\sigma_x&, \mu=y \\
	0 &, \mu=z \\       
\end{array}
\right .
\end{equation}
Therefore, the hadronic tensor $W_{\mu\nu}$ can be written as follows:
$$W_{\mu\nu}=\overline{(\chi_2^\dagger F_{\mu}\chi_1)(\chi_1^\dagger F_{\nu}^\dagger\chi_2)}=\displaystyle\frac{1}{2} TrF_{\mu}\rho_1
F_{\nu}^\dagger\rho_2; $$
where the averaging (summing) acts only on the two-component spinors, and we introduced density matrix for the nucleon: $\rho=\chi\chi^\dagger$, $\rho_{ab}=\chi_a\chi_b^*$, and $a,b=1,2$ are the spinor indexes. We included the statistical factor $1/(2s+1)=1/2$, for the initial nucleon.

In case of unpolarized particles $\rho=1/2$, and $$W_{\mu\nu}=\displaystyle\frac{1}{2} TrF_{\mu}F_{\nu}^\dagger. $$

\subsection{Leptonic tensor $L_{\mu\nu}$ }

The leptonic tensor, which describes the electron vertex, is written as:
$$L_{\mu\nu}=\overline{\ell_{\mu}\ell_{\nu}^*}=\overline{\overline{u}(k_2)\gamma_{\mu}u(k_1)\left [\overline{u}(k_2)\gamma_{\nu}u(k_1)\right ]^*}.$$
where the overline denotes the averaging over the polarizations of the initial electron and the summing over the polarizations of the final electrons. Recalling that 
$$\overline{u}=u^\dagger \gamma_0,~~\overline{u}^\dagger=(u^\dagger\gamma_0)^\dagger=\gamma_0^\dagger u=\gamma_0u,~~\gamma_0\gamma_0=1,~\gamma_0^\dagger=\gamma_0,
$$
we can write:
\begin{equation}
\begin{array}{ll}
L_{\mu\nu}&=\overline{\overline{u}(k_2)\gamma_{\mu}u(k_1)
u^\dagger(k_1)\gamma_{\nu}^\dagger\overline{u}(k_2)}
=\overline{\overline{u}(k_2)\gamma_{\mu}u(k_1)
u^\dagger(k_1)\gamma_0
\gamma_0\gamma_{\nu}^\dagger\gamma_0 u(k_2)}\\
&=\overline{\overline{u}(k_2)\gamma_{\mu}u(k_1)
u^\dagger(k_1)\gamma_{\nu}^\dagger\gamma_0 u(k_2)}=
\displaystyle\frac{1}{2}Tr \gamma_{\mu}\rho_e^1\gamma_{\nu}\rho_e^2.
\label{eq:eql}
\end{array}
\end{equation}

From the Dirac theory we can write: $\overline{\overline{u}(k)u^\dagger (k)}=\hat k+m_e=\rho$. After performing the corresponding substitutions in Eq. (\ref{eq:eql}), one finds (see Appendix 2):
$$L_{\mu\nu}=\displaystyle\frac{1}{2}Tr \gamma_{\mu}(\hat{k_1}+m_e)\gamma_{\nu}(\hat{k_2}+m_e),$$
from where we derive (neglecting the electron mass):
\begin{center}
\fbox{$
L_{\mu\nu}=2k_{1\mu}k_{2\nu}+2k_{1\nu}k_{2\mu}-2g_{\mu\nu}k_1\cdot k_2$.}
\end{center}

From this expression we see that the leptonic tensor which describes unpolarized electrons is symmetrical.

\subsection{The Rosenbluth formula}

Let us calculate explicitly the components for the hadronic tensor $W_{\mu\nu}$, in terms of the FFs $G_{EN}$ and $G_{MN}$. Recalling the property that $Tr \vec\sigma\cdot \vA=0$, for any vector $\vA$, we see that all terms for the components $W_{\mu\nu}$ which  contain the product $G_{EN}G_{MN}$ vanish: this means that the unpolarized cross section of $eN-$scattering does not contain this interference term. The non-zero components of $W_{\mu\nu}$ are determined only by $G_{EN}^2$ and $G_{MN}^2$:
$$W_{00}=4m^2G_{EN}^2,$$
$$W_{xx}=-q^2G_{MN}^2,$$
$$W_{yy}=-q^2G_{MN}^2.$$ 
Substituting these expressions in Eq. (\ref{eq:cscm}), one can find for the matrix element squared:
\begin{equation}
\left (\displaystyle\frac{q^2}{e^2}\right )^2\overline{\left| {\cal M}\right |^2}
=L_{00}W_{00}+(L_{xx}+L_{yy})W_{xx}=L_{00}4m^2G_{EN}^2+(L_{xx}+L_{yy})(-q^2)G_{MN}^2.
\label{eq:msq}
\end{equation}
The necessary components of the leptonic tensor $L_{\mu\nu}$, calculated in the Breit system, are:
$$L_{00}=4\epsilon_{1B}^2+q^2=-q^2\cot^2\displaystyle\frac{\theta_B}{2},$$
$$L_{yy}=-q^2,$$
$$L_{xx}=4k_{1x}^2-q^2=-q^2\left (1+\cot^2\displaystyle\frac{\theta_B}{2} \right ).$$
Substituting the corresponding terms in Eq. (\ref{eq:msq}) we have:
$$
\overline{\left |{\cal  M}\right |^2}=\left (\displaystyle\frac {e^2}{q^2}\right )^2
\left [ -q^2\cot^2 \displaystyle\frac{\theta_B}{2}4m^2 G_{EN}^2+(-q^2-q^2\cot^2\displaystyle\frac{\theta_B}{2})(-q^2G_{MN}^2)\right ],
$$
which becomes in the Lab system:
\begin{equation}
\overline{\left |{\cal  M}\right |^2}=\left (\displaystyle\frac{e^2}{q^2}\right )^2 4m^2(-q^2)\left [2\tau G_{MN}^2+
\displaystyle\frac{\cot^2\frac{\theta_e}{2}}{1+\tau}(G_{EN}^2+\tau G_{MN}^2)\right ].
\label{eq:msqa}
\end{equation}
We can then find the following formula for the cross section, ${d\sigma}/{d\Omega}_e$, in the Lab system, in terms of the electromagnetic FFs $G_{EN}$ and $G_{MN}$ (Rosenbluth formula \cite{Rose}) \footnote{More exactly, the original formula has been written in terms of the Dirac ($F_1$) and Pauli ($F_2$) form factors.}:
\begin{equation}
\displaystyle\frac{d\sigma}{d\Omega}_e=\displaystyle\frac{\alpha^2}{-q^2}
\left (\displaystyle\frac{\ed}{\eu}\right )^2 \left [2\tau G_{MN}^2+
\displaystyle\frac{\cot^2\frac{\theta_e}{2}}{1+\tau}\left (G_{EN}^2+\tau G_{MN}^2\right )\right ],
\label{eq:csmf}
\end{equation}
where $\alpha=e^2/4\pi\simeq 1/137$ is the fine structure constant. 

Taking into account Eq. (\ref{eq:eq8}) and the following relation between the energy $\ed$ and the angle $\theta_e$ of the scattered electron:
\begin{equation}
\ed=\displaystyle\frac{\eu}{1+2\displaystyle\frac{\eu}{m}\sin^2
\displaystyle\frac{\theta_e}{2}},
\label{eq:e1e2}
\end{equation}
the differential cross section can be written in the following form:
\begin{equation}
\displaystyle\frac{d\sigma}{d\Omega}_e=\sigma_M\left [2\tau G_{MN}^2\tan^2\frac{\theta_e}{2} +
\displaystyle\frac{G_{EN}^2+\tau G_{MN}^2}{1+\tau}\right ],
\label{eq:csst}
\end{equation}
with 
$$\sigma_M=
\displaystyle\frac{\alpha^2}{-q^2}
\left (\displaystyle\frac{\ed}{\eu}\right )^2\displaystyle\frac{\cos^2
\displaystyle\frac{\theta_e}{2}}{\sin^2
\displaystyle\frac{\theta_e}{2}}= \left (\displaystyle\frac{\alpha}{2\eu}\right )^2\displaystyle\frac{\cos^2
\displaystyle\frac{\theta_e}{2}}{\sin^4
\displaystyle\frac{\theta_e}{2}} \displaystyle\frac{1}{\left (1+2\displaystyle\frac{\eu}{m}\sin^2
\displaystyle\frac{\theta_e}{2}\right )}$$
where $\sigma_M$ is the so called Mott cross section, for the scattering of unpolarized electrons by a point charge particle (with spin 1/2).

Note that the very specific $\cot^2\displaystyle\frac{\theta_e}{2}$-dependence of the cross section for $eN$-scattering results from the assumption of one-photon mechanism for the considered reaction.

This can be easily proved \cite{Re99}, by cross-symmetry considerations, looking to the annihilation channel, $e^++e^-\to p+\overline{p}$. In the CMS of such reaction, the one-photon mechanism induces a simple and evident $\cos^2\theta$-dependence of the corresponding differential cross section, due to the C-invariance of the hadron electromagnetic interaction, and unit value of the photon spin.

\begin{center}
\begin{figure}[h]
\hspace*{3true cm}
\mbox{\epsfxsize=10.cm\leavevmode\epsffile{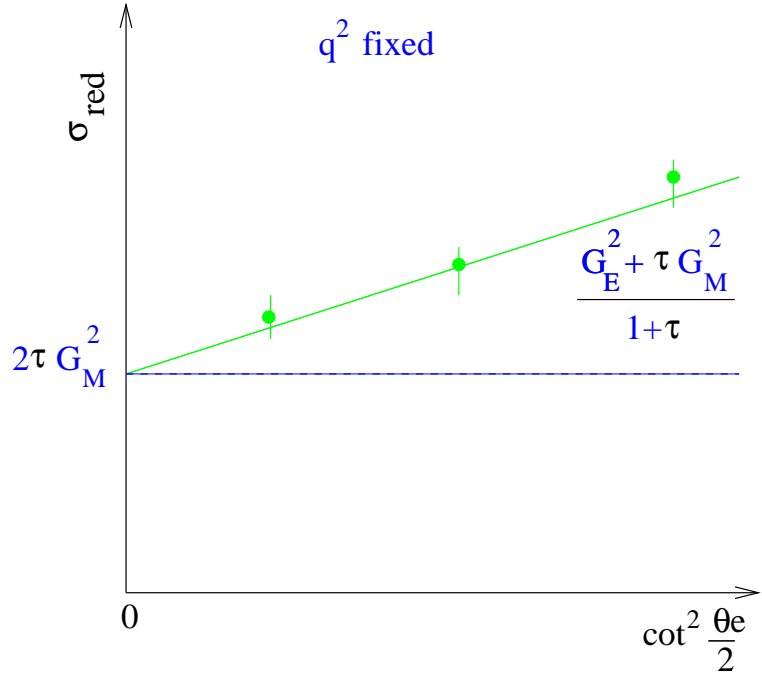}}
\caption{Illustration of the Rosenbluth separation for the elastic differential cross section for $eN$-scattering.}
\label{fig:ros}
\end{figure}
\end{center}

The particular $\cot^2\displaystyle\frac{\theta_e}{2}$-dependence of the differential $eN$-cross section is at the basis of the method to determine both nucleon electromagnetic FFs, $G_{EN}$ and $G_{MN}$, using the linearity of the {\it reduced} cross section:
$$\sigma_{red}=\displaystyle\frac{\displaystyle\frac{d\sigma}{d\Omega}_e}
{\displaystyle\frac{\alpha^2}{-q^2}\left(\displaystyle\frac{\ed}{\eu}\right )^2}$$
as a function of $\cot^2\frac{\theta_e}{2}$ (Rosenbluth fit or Rosenbluth separation). One can see that the backward $eN$-scattering ($\theta_e=\pi,\cot^2\frac{\theta_e}{2}=0$) is determined by the magnetic FF only, and that the slope for $\sigma_{red}$ is sensitive to $G_{EN}^2$ (Fig. \ref{fig:ros}).
\vspace*{-1 true cm}

At large $q^2$, (such that $\tau\gg 1$), the differential cross-section ${d\sigma}/{d\Omega}_e$ (with unpolarized particles) is insensitive to $G_{EN}$: the corresponding combination of the nucleon FFs, $G_{EN}^2+\tau G_{MN}^2$ is dominated by the $G_{MN}$ contribution, due to the following reasons:
\begin{itemize}
\item $G_{Mp}/G_{Ep}\simeq \mu_p$, where $\mu_p$ is the proton magnetic moment, so $G_{Mp}^2/G_{Ep}^2\simeq 2.79^2\simeq 8$;
\item The factor $\tau$ increases the $G_{MN}^2$ contribution at large momentum transfer, where $\tau\gg 1$.
\end{itemize}
Therefore $ep-$scattering (with unpolarized particles) is dominated by the magnetic FF, at large values of momentum transfer. The same holds for $en-$scattering, even at relatively small values of $q^2$, due to the smaller values of the neutron electric FF. 

As a result, for the exact determination of the proton electric FF, in the region of large momentum transfer,  and for the neutron electric FF - at any value of $q^2$, polarization measurements are required and in particular those polarization observables which are determined by the product $G_{EN}G_{MN}$, and are, therefore, more sensitive to $G_{EN}$.

There are at least two different classes of polarization experiments of such type: the scattering of longitudinally polarized electrons by polarized target (with polarization in the reaction plane, but perpendicular to the direction of the three-momentum transfer) $\vec e +\vec p\to e+p$, or the measurement of the ratio of transversal to longitudinal proton polarization (in the reaction plane) for the scattering of longitudinally polarized electrons by unpolarized target, $\vec e + p\to e+\vec p$.

In principle, there are some components of the depolarization tensor (characterizing the dependence of the final proton polarization on the target polarization (for the scattering of unpolarized electrons, $e + \vec p\to e+\vec p$) which are also proportional to $G_{EN}G_{MN}$, and therefore can be used for the determination of the nucleon electric FF \cite{Re68,Do69w}.

Both experiments (with polarized electron beam) have been realized: $p(\vec e, \vec p) e$ for the determination of $G_{Ep}$ \cite{Jo00} and, for the determination of $G_{En}$, $\vec d(\vec e, e'n) p$  \cite{Gen1} and $d(\vec e, e'\vec n) p$ \cite{Gen2}.

\section{Polarization observables}

In general the hadronic tensor $W_{\mu\nu}$, for $ep$ elastic scattering, contains four terms, related to the 4 possibilities of polarizing the initial and final protons:
$$W_{\mu\nu}=W_{\mu\nu}^{(0)}+W_{\mu\nu}(\vPu)+W_{\mu\nu}(\vPd)+W_{\mu\nu}(\vPu,\vPd),$$
where $\vPu~(\vPd)$ is the polarization vector of the initial (final) proton.
The first term corresponds to the unpolarized case, the second (third) term corresponds to the case when the initial (final) proton is polarized, and the last term describes the reaction when both protons (initial and final) are polarized. The $2\times 2$ density matrix for a nucleon with polarization $\vP$ can be  written as:
$\rho=\displaystyle\frac{1}{2}\left (1+\vec \sigma\cdot\vP\right )$. 

Let us consider the case when only the final proton is polarized 
($\vP=\vPd$):
$$W_{\mu\nu}(\vP)=\displaystyle\frac{1}{2}Tr F_{\mu}F_{\nu}^\dagger
\vec \sigma\cdot\vP
$$
For the scattering of longitudinally polarized electrons (by unpolarized target), only the $x$ and $z$ components of the polarization vector $\vP$ do not vanish. To find these components, let us calculate the tensors $W_{\mu\nu}(P_x)$ and $W_{\mu\nu}(P_z)$.
$$W_{\mu\nu}(P_x)=\displaystyle\frac{1}{2}Tr F_{\mu}F_{\nu}^\dagger \sigma_x.$$
Let us start \footnote{ We will take into account the fact that the FFs $G_{EN}(q^2)$ and $G_{MN}(q^2)$ are real functions of $(q^2)$ in the space-like region, see later.}
from the calculation of the components $F_{\nu}^\dagger$: 

\begin{equation}
F_{\nu}^\dagger =\left\{ \begin{array}{ll}
         2mG_{EN}& ,\nu=0 \\
         -i\sqrt{-q^2} G_{MN}\sigma_y&,\nu=x \\
     i\sqrt{-q^2} G_{MN}\sigma_x&,\nu=y \\
	0 &,\nu=z \\       
\end{array}
\right .
\end{equation}
Therefore, one can find easily (using $\sigma_x\sigma_y=i\sigma_z$,
$\sigma_y\sigma_z=i\sigma_x$,$\sigma_z\sigma_x=i\sigma_y$):
\begin{equation}
F_{\nu}^\dagger \sigma_x=\left\{ \begin{array}{ll}
         2mG_{EN}\sigma_x& ,\nu=0 \\
         -\sqrt{-q^2} G_{MN}\sigma_z&,\nu=x \\
      i\sqrt{-q^2} G_{MN}&,\nu=y \\
	0 &,\nu=z. \\       
\end{array}
\right .
\end{equation}
This allows to write:
\begin{equation}
F_{\mu}F_{\nu}^\dagger \sigma_x =\left\{ \begin{array}{ll}
         2mG_{EN}& ,\mu=0 \\
         i\sqrt{-q^2} G_{MN}\sigma_y&,\mu=x \\
     -i\sqrt{-q^2} G_{MN}\sigma_x&,\mu=y \\
	0 &,\mu=z \\       
\end{array}
\right .
~\bigotimes~\left\{ \begin{array}{ll}
         2mG_{EN}\sigma_x& ,\nu=0 \\
         -\sqrt{-q^2} G_{MN}\sigma_z&,\nu=x \\
         i\sqrt{-q^2} G_{MN}&,\nu=y \\
	0 &,\nu=z \\       
\end{array}
\right .
\end{equation}
As we have to calculate the trace, recalling that $Tr \sigma_{x,y,z}=0$,
we can see that the non-zero components of the hadronic tensor $W_{\mu\nu}(P_x)$ are:
\begin{equation}
\begin{array}{ll}
W_{0y}(P_x)=&i\sqrt{-q^2} ~2mG_{EN}G_{MN},\\
W_{y0}(P_x)=&-i\sqrt{-q^2} ~2mG_{EN}G_{MN}.\\
\end{array}
\end{equation}

\begin{figure}[h]
\mbox{\epsfxsize=14.cm\leavevmode\epsffile{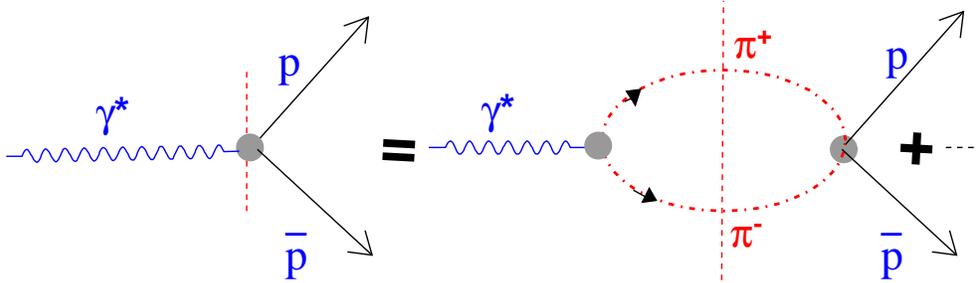}}
\caption{The unitarity condition for proton electromagnetic FFs in the time-like region of momentum transfer squared. Vertical line on the right side crosses the pion lines, describing real particles (on mass shell). The dotted line denotes other possible multi-pion states, in the chain of the following transitions: $\gamma^*\to n\pi\to p\overline{p}$, where $n$ is the number of pions in the intermediate state.}
\label{fig:uni}
\end{figure}

So we proved here that only two components of $W_{\mu\nu}(P_x)$ are different from zero: they are equal in absolute value and opposite in sign: it follows that $W_{\mu\nu}(P_x)$ is an antisymmetrical tensor.
Therefore, the product $L_{\mu\nu}W_{\mu\nu}(P_x)$ vanishes: $L_{\mu\nu}$ is a symmetrical tensor: the product of a symmetrical tensor and an asymmetrical tensor is zero. This means that the polarization of the final proton vanishes, if the electron is unpolarized: {\bf unpolarized electrons can not induce polarization of the scattered proton}. This is a property of the one-photon mechanism $for~any~elastic~electron-hadron~scattering$ and of the hermiticity of the Hamiltonian for the hadron electromagnetic interaction. Namely the hermiticity condition allows to prove that the hadron electromagnetic FFs are real functions of the momentum transfer squared in the space-like region. On the other hand, in the time-like region, which is scanned by the annihilation processes, $e^-+e^+\leftrightarrow p+\overline{p}$, the nucleon electromagnetic FFs are complex functions of $q^2$, if $q^2\ge 4m_\pi^2$, where $m_\pi$ is the pion mass. This is due to the unitarity condition, which can be illustrated as in Fig. \ref{fig:uni}.
 
The complexity of nucleon FF's (in the time-like region) results in specific polarization phenomena, for the the annihilation processes $e+e^-\leftrightarrow p+\overline{p}$, which are different from the case of elastic $ep-$scattering. For example, the polarization of the final proton (or antiproton) is different from zero, even in the case of collisions of unpolarized leptons: this polarization is determined by the product ${\cal I}m G_{EN}G_{MN}^*$ (and, therefore vanishes in the case of elastic $ep$-scattering, where the FFs are real). 
Note that two-photon exchange in $ep$-elastic scattering is also generating complex amplitudes. So the interference between one and two-photon amplitudes induces nonzero proton polarization, but small in absolute value, as it is proportional to $\alpha$.

Numerous experiments \cite{exp} have been done with the aim to detect such polarization at small momentum transfer $|q^2|\le$ 1 GeV$^2$, but with negative result, at a percent level.
Only recently the above mentioned interference was experimentally detected, measuring the asymmetry in the scattering of transversally polarized electrons by an unpolarized proton target \cite{We01}.

Note that at very large momentum transfer, the relative role of two-photon amplitudes may be increased (violating the counting in $\alpha$), due to the steep $q^2$-decreasing of hadronic electromagnetic FFs.

Note also that the analytical properties of the nucleon FFs, considered as functions of the complex variable $z=q^2$, result in a specific asymptotic behavior, as they obey to the Phragm\`en-Lindel\"of theorem \cite{Ti39}:
\begin{equation}
\lim_{q^2\to -\infty} F^{(SL)}(q^2) =\lim_{q^2\to \infty} 
F^{(TL)}(q^2).
\label{eq:eqpl}
\end{equation}
The existing experimental data about the proton FFs in the time-like region up to 15 GeV$^2$, seem to contradict this theorem \cite{Re01}. More exactly, one can prove that, if one FF, electric or magnetic; satisfies the relation (\ref{eq:eqpl}), then the other one violates this theorem, i.e. the asymptotic condition does not apply.

Let us consider now the proton polarization in the $z$-direction:
$$W_{\mu\nu}(P_z)=\displaystyle\frac{1}{2}Tr F_{\mu}F_{\nu}^\dagger \sigma_z.$$
Firstly we calculate the components of $F_{\nu}^\dagger \sigma_z$:
\begin{equation}
F_{\nu}^\dagger \sigma_z= \left\{ \begin{array}{ll}
         2mG_{EN}\sigma_z& ,\nu=0 \\
         \sqrt{-q^2} G_{MN}\sigma_x&,\nu=x \\
     \sqrt{-q^2} G_{MN}\sigma_y&,\nu=y \\
	0 &,\nu=z \\       
\end{array}
\right .
\end{equation}
Therefore we find:
\begin{equation}
F_{\mu}F_{\nu}^\dagger \sigma_z =\left\{ \begin{array}{ll}
         2mG_{EN}& ,\mu=0 \\
         i\sqrt{-q^2} G_{MN}\sigma_y&,\mu=x \\
     -i\sqrt{-q^2} G_{MN}\sigma_x&,\mu=y \\
	0 &,\mu=z \\       
\end{array}
\right .
\bigotimes\left\{ \begin{array}{ll}
         2mG_{EN}\sigma_z&, \nu=0 \\
         \sqrt{-q^2} G_{MN}\sigma_x&,\nu=x \\
       \sqrt{-q^2} G_{MN}\sigma_y&,\nu=y \\
	0 &,\nu=z \\       
\end{array}
\right .
\end{equation}
We see that $W_{0\nu}(P_z)=W_{\nu 0}(P_z)=0$, for any $\nu$, and no interference term $G_{EN}G_{MN}$ is present. The nonzero components of $W_{\mu\nu}(P_z)$ are:
\begin{equation}
 \begin{array}{ll}
W_{xy}(P_z)=&-iq^2 G_{MN}^2,\\
W_{yx}(P_z)=&iq^2 G_{MN}^2,\\
  \end{array}
\label{eq:pz0}
\end{equation}
from where we see that $W_{\mu\nu}(P_z)$ is an antisymmetrical tensor, which depends on $G_{MN}^2$ and  that $P_x/P_z \propto G_{EN}/G_{MN}$.

\subsection{Polarized electron}
The leptonic tensor, $L_{\mu\nu}$, in case of unpolarized particles, contains only one term. For longitudinally polarized electrons, the polarization is characterized by the helicity $\lambda$, which takes values $\pm 1$, corresponding to the direction of spin parallel or antiparallel to the electron three-momentum.
The general expression for the leptonic tensor is:
\begin{equation}
L_{\mu\nu}=L_{\mu\nu}^{(0)}+L_{\mu\nu}(\lambda_1)+L_{\mu\nu}(\lambda_2)+
L_{\mu\nu}(\lambda_1,\lambda_2).
\label{eq:lepg}
\end{equation}
where the first term, considered previously, describes the collision where the initial and final electrons are unpolarized, the second (third) term describes the case when the initial (final) electron is longitudinally polarized, and the last terms holds when both electrons are polarized.
If only the initial electron is polarized, $\lambda_1=\lambda$, one can write for $L_{\mu\nu}$ (see Appendix II):
\begin{equation}
L_{\mu\nu}(\lambda)=2i\lambda\epsilon_{\mu\nu\alpha\beta}k_{1\alpha}k_{2\beta}.
\label{eq:lep}
\end{equation}

The effect of the electron polarization is described by an antisymmetrical tensor $L_{\mu\nu}(\lambda)$. If the initial proton is unpolarized, again, being described by symmetrical tensor, the total result will be zero. This result holds because the FFs are real, so it does not apply in the time-like region.

\vspace{.2true cm}

\noindent\underline{$x$-component} 

Let us consider the product of the leptonic $L_{\mu\nu}(\lambda)$ and hadronic 
$W_{\mu\nu}(P_x)$ tensors, for the $x$ component of the final proton polarization:
\begin{equation}
\begin{array}{ll}
L_{\mu\nu}(\lambda)W_{\mu\nu}(P_x)&=L_{0y}(\lambda)W_{0y}(P_x)+L_{y0}(\lambda)W_{y0}(P_x) \\
&=L_{0y}(\lambda)\left [W_{0y}(P_x)-W_{y0}(P_x)\right ]=2L_{0y}(\lambda) W_{0y}(P_x).\\
\end{array}\label{eq:pxl}
\end{equation}
Taking into account that: $L_{0y}=2i\lambda\epsilon_{0y\alpha\beta}k_{1\alpha}k_{2\beta}$ the only non-zero terms correspond to $\alpha=x$ and $\beta=z$ or $\alpha=z$ and $\beta=x$.
Therefore:
$$L_{0y}(\lambda) = 2i\lambda\left (\epsilon_{0yxz}k_{1x}k_{2z}+\epsilon_{0yzx}k_{1z}k_{2x}\right )
=2i\lambda\epsilon_{0yxz}(k_{1x}k_{2z}-k_{1z}k_{2x})=
i\lambda q^2 \cot \displaystyle\frac{\theta_B}{2} ,$$
with $\epsilon_{0yxz}=1$, and using Eqs. (\ref{eq:vkube}) and (\ref{eq:vkdbe}).

We finally find:
\begin{equation}
L_{\mu\nu}(\lambda)W_{\mu\nu}(P_x)=-4\lambda m q^2\sqrt{-q^2}\cot \displaystyle\frac{\theta_B}{2}G_{EN}G_{MN}.
\label{eq:pe}
\end{equation}

\noindent\underline{$z$-component}

Similarly, considering the antisymmetry of both tensors 
$L_{\mu\nu}(\lambda)$ and $W_{\mu\nu}(P_z)$, one can find:
\begin{eqnarray}
L_{\mu\nu}(\lambda)W_{\mu\nu}(P_z)&=2i\lambda\epsilon_{\mu\nu\alpha\beta}
k_{1\alpha}k_{2\beta}W_{\mu\nu}(P_z)=4\epsilon_{xy0z}W_{xy}(P_z)\left (\eub k_{2B}^z-\edb k_{1B}^z\right )\nonumber\\
&= 4\lambda q^2\displaystyle\frac{G_{MN}^2}{\sin\theta_B/2}.
\label{eq:pz1}
\end{eqnarray}

\section{Final formulas}

The polarization $\vP$ of the scattered proton can be written as:

$$\vP\displaystyle\frac{d\sigma}{d\Omega_e}=
\displaystyle\frac{\alpha^2}{4\pi^2}
\left (\displaystyle\frac{\ed}{\eu}\right )^2
\displaystyle\frac{L_{\mu\nu}}{m^2}
\vec P_{\mu\nu}.$$
with 
$\vec P_{\mu\nu}=\displaystyle\frac{1}{2}({\cal T}r{\cal F}_{\mu} {\cal F}_{\nu}^\dagger\vec\sigma)$, so that $P_{\mu\nu}^{(z)}= W_{\mu\nu}(P_z)$ and 
$P_{\mu\nu}^{(x)}= W_{\mu\nu}(P_x)$

Using Eq. (\ref{eq:cot}) one can find the following expressions for the components 
$P_x$ and $P_z$ of the proton polarization vector (in the scattering plane) - in terms of the proton electromagnetic FFs:
\begin{equation}
\begin{array}{ll}
DP_x&=-2\lambda \cot \displaystyle\frac{\theta_e}{2} \sqrt{\displaystyle\frac{\tau}{1+\tau }}G_{EN}G_{MN},\\
DP_z&=\lambda\displaystyle\frac{\eu+\ed}{m}\sqrt{\displaystyle\frac{\tau}{1+\tau }}G_{MN}^2,\\
\end{array}
\label{eq:fi}
\end{equation}
where D is proportional to the differential cross section with unpolarized particles:
\begin{equation}
D=2\tau G_{MN}^2+\cot^2 \displaystyle\frac{\theta_e}{2} \displaystyle\frac{G_{EN}^2+\tau G_{MN}^2}{1+\tau }.
\label{eq:mott}
\end{equation}
So, for the ratio of these components one can find the following formula:
\begin{equation}
\displaystyle\frac{P_x}{P_z}=\displaystyle\frac{P_t}{P_\ell}= - 2\cot \displaystyle\frac{\theta_e}{2} \displaystyle\frac{m}{\eu+\ed}\displaystyle\frac{G_{EN}(q^2)}{G_{MN}(q^2)}
\label{eq:final}
\end{equation}
which clearly shows that a  measurement of the ratio of transverse and longitudinal polarization of the recoil proton gives is a direct measurement of the ratio of electric and magnetic FFs, $G_{EN}(q^2)/G_{MN}(q^2)$.

In the same way it is possible to calculate the dependence of the differential cross section for the elastic scattering of the longitudinally polarized electrons by a {\bf polarized} proton target, with polarization ${\cal P}$, in the above defined coordinate system:
\begin{equation}
\displaystyle\frac{d\sigma}{d\Omega_e}({\cal P})=
\left (\displaystyle\frac{d\sigma}{d\Omega_e}\right )_0 
\left ( 1+\lambda {\cal P}_xA_x+\lambda {\cal P}_zA_z\right ),
\label{eq:tpol}
\end{equation}
where the asymmetries $A_x$ and $A_z$ (or the corresponding analyzing powers) are related in a simple and direct way, to the components of the final proton polarization:
\begin{equation}
\begin{array}{ll}
A_x&=P_x,\\
A_z&=-P_z.\\
\end{array}
\label{eq:ap}
\end{equation}
This holds in the framework of the one-photon mechanism for elastic $ep-$scattering. Note that the quantities 
$A_x$ and $P_x$ have the same sign and absolute value, but the components $A_z$ and $P_z$, being equal in absolute value, have opposite sign.

These two different polarization experiments in elastic electron-proton scattering, namely the scattering with longitudinally polarized electrons by a polarized proton target (with polarization in the reaction plane) from one side and the measurement of the components of the final proton polarization (again in the reaction plane)
in the scattering of longitudinally polarized electrons by an unpolarized proton target, from another side, bring the same physical information, concerning the electromagnetic FFs of proton.

Note that the $P_y$-component of the proton polarization vanishes in the scattering of polarized and unpolarized electrons, as well. This results from the one-photon mechanism and the reality of form factors $G_{EN}$ and $G_{MN}$. For the same reasons, the corresponding analyzing power, $A_y$, also vanishes.

Summarizing this discussion, let us stress once more that these results for polarization observables in elastic $ep$-scattering hold in the framework of the one-photon mechanism. 

Still in the framework of the one-photon mechanism, there are at least two different sources of corrections to these relations:
\begin{itemize}
\item the standard radiative corrections;
\item the electroweak corrections.
\end{itemize}

These last corrections arise from the interference of amplitudes, corresponding to the exchange of $\gamma$ and $Z-$boson. The relative value of these contributions is characterized by the following dimensionless parameter:
$$
G_{eff}=\displaystyle\frac{G_F}{2\sqrt{2}\alpha\pi} 
|q^2|\simeq 10^{-4} \displaystyle\frac{|q^2|}{\mbox{GeV}^2},
$$ 
where $G_F$ is the standard Fermi constant of the weak interaction, $G_F\simeq$ 10 $^{-5}/m^2$.

\begin{figure}[h]
\hspace*{3true cm}
\mbox{\epsfxsize=7.cm\leavevmode\epsffile{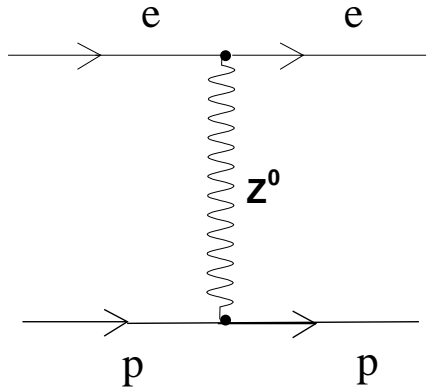}}
\caption{ Feynman diagram corresponding to $Z$-exchange, in $ep$-scattering}
\label{fig:zbs}
\end{figure}

So, for $|q^2|\le$ 10 GeV$^2$, the electroweak corrections are negligible, for the polarization phenomena considered above. However, note that the $\gamma\bigotimes Z$-interference is not only inducing (small) corrections to the results of the one-photon considerations, but it induces also a new class of polarization observables of P-odd nature, i.e. with violation of the $P$-invariance. The simplest of them is the P-odd asymmetry of the scattering of longitudinally polarized electrons by an unpolarized proton target $\vec e+p\to e+p$ (the detection of the polarization of the scattered particles is not required).  As this asymmetry vanishes in the one-photon mechanism, it is proportional to $G_{eff}$, at relatively small momentum transfer squared.

Let us turn to the QED radiative corrections. They appear essentially in  the differential cross section, and they have been discussed in particular in \cite{Mo69} and more recently by \cite{Ma99,Ma00}

For polarization phenomena, it can be proved \cite{Ma00} that, in case of soft photons, the contribution of radiative corrections can be explicitly factorized. Therefore this contribution, which is important for the differential cross section, cancels in polarization effects. Radiation of non-soft photons by electrons (in initial and final states) results in corrections, which are different for the components $P_x$ and $P_z$. Such corrections can be calculated in a model independent way, in the framework of the standard QED, inducing effects of a few percent \cite{Af01,Afa01}. 

\begin{figure}[h]
\hspace*{3true cm}
\mbox{\epsfxsize=5.cm\leavevmode\epsffile{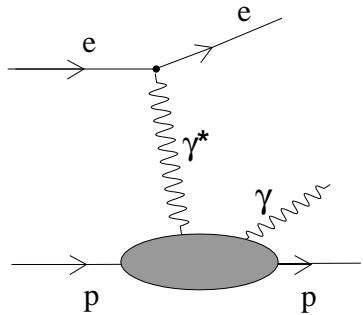}}
\caption{Feynman diagram corresponding to virtual Compton scattering}
\label{fig:vcs}
\end{figure}

\begin{center}
\begin{figure}
\mbox{\epsfxsize=14.cm\leavevmode\epsffile{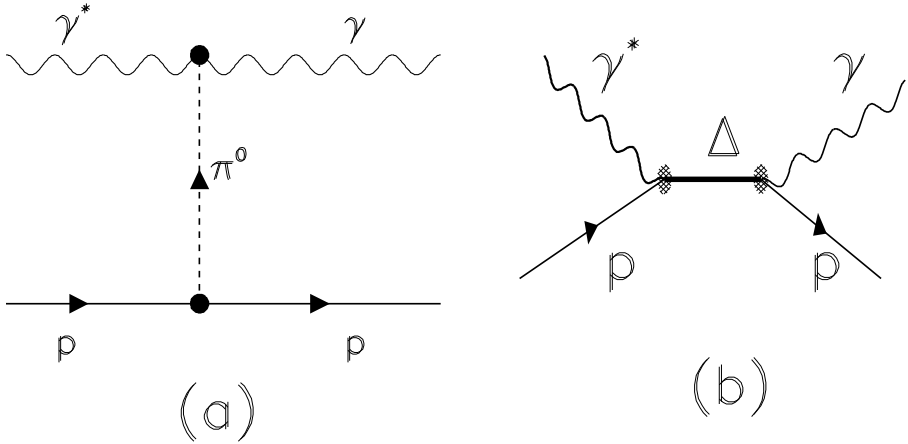}}
\caption{Feynman diagrams for $\gamma^*+ p\to\gamma+ p$: one-pion exchange in $t$-channel (a);
$\Delta$-exchange in $s$-channel (b).}
\label{fig:vcm}
\end{figure}
\end{center}

Model dependent radiative corrections can not be uniquely calculated. This concerns, first of all, the virtual Compton scattering on nucleons (Fig. \ref{fig:vcs}), which is driven by the amplitude of the process $\gamma^*+p\to \gamma +p$- with very complicated spin structure and with different mechanisms, as, for example, pion exchange in $t$-channel and $\delta$-exchange in $s$-channel (Fig. \ref{fig:vcm}). These contributions can be estimated to give corrections of 1-3 \%.

The interference amplitude between the Bethe-Heitler mechanism with the VCS amplitude has to be more important, as it can not be factorized, even for soft photon radiation. However the largest contribution has to be due to the two-photon exchange mechanism, if present \cite{Gu73}, when the (large) momentum transfer is equally shared between the two virtual photons .

\section{Concluding remarks}

We reproduced above the scheme for the calculation of the simplest non-zero polarization phenomena for elastic $eN$-scattering, in the framework of the one-photon mechanism, in terms of the nucleon electromagnetic form factors $G_{EN}$ and $G_{MN}$. We showed the interest of the Breit system, where all calculations are simplified and the different terms have a transparent physical interpretation. Such calculations are done in terms of measurable quantities, such as the nucleon polarization vector, in the system where the nucleon 
is in rest.

This technique can be easily generalized to the case of P-odd polarization phenomena in $eN$-scattering, induced by the interference of $\gamma$ and $Z$-exchanges, in the framework of the Standard Model. These calculations can be done in a model independent way.

We only mention such complicated problems as radiative corrections. Only for Bethe-Heitler bremsstrahlung polarization phenomena can be derived in model independent form, in terms of nucleon electromagnetic FFs. But virtual Compton scattering is essentially model dependent.  

The most intriguing part of the radiative corrections is due to the two-photon exchange at large momentum transfer, with comparable virtuality of the two photons, for which no calculation actually exists.

This formalism equally applies to $en$-elastic scattering, too, in the case of free neutron. As typically a target like $d$ or $^3\!He$ is used, specific considerations apply, which are outside the present notes. This formalism is valid in case of elastic $e+^3\!He$ and $e+^3\!H$ scattering, and, in general, for elastic scattering of electrons on any spin 1/2 target.

 Polarization phenomena for elastic positron scattering and for elastic scattering of positive and negative muons are the same as in case of electron scattering.

\section{Appendix I: Pauli term}

We derive the following relation: 
\begin{equation}
\overline{u}(p_2)
\displaystyle\frac{\sigma_{\mu\nu}q_\nu}{2m}u(p_1)
=\overline{u}(p_2)\left [\gamma_{\mu}-\displaystyle\frac{(p_1+p_2)_\mu}{2m}\right ]u(p_1).
\label{eq:apr}
\end{equation}
Using the definition for $\sigma_{\mu\nu}$, one can write:
$$\overline{u}(p_2)\displaystyle\frac{\gamma_{\mu}\gamma_{\nu}-\gamma_{\nu}\gamma_{\mu}}
{4m}q_{\nu}u(p_1)=\overline{u}(p_2)\displaystyle\frac{\gamma_{\mu}\hat q-\hat q\gamma_{\mu}}{4m}u(p_1).$$

Recalling that $q=p_2-p_1$ with $\hat a=a_\mu\gamma_{\mu}$:
$$\overline{u}(p_2)\displaystyle\frac{\gamma_{\mu}(\hat{p_2}-\hat{p_1})
-(\hat{p_2}-\hat{p_1})\gamma_{\mu}}{4m}u(p_1).$$
Applying the Dirac equation:
$$(\hat p-m)u(p)=0 \rightarrow \hat p u(p)=mu(p),$$
$$\overline{u}(p)(\hat p-m)=0 \rightarrow  \overline{u}(p)\hat p=\overline{u}(p)m,$$
we find:
\begin{equation}
\overline{u}(p_2)\displaystyle\frac{\gamma_{\mu}(\hat{p_2}-m)-(m-\hat{p_1})
\gamma_{\mu}}{4m}u(p_1)
=-\displaystyle\frac{1}{2}\overline{u}(p_2)\gamma_{\mu}u(p_1)+
\displaystyle\frac{1}{4m}\overline{u}(p_2)\left [\gamma_{\mu}\hat{p_2}+\hat{p_1}\gamma_{\mu} \right ]u(p_1)
\label{eq:ap1}
\end{equation}
Using the properties: $\gamma_{\mu}\gamma_{\nu}+\gamma_{\nu}\gamma_{\mu}=2g_{\mu\nu}$, $\hat a \hat b+\hat b\hat a=2ab$, $\hat a\gamma_{\mu}+\gamma_{\mu}\hat a=2
a_{\mu}$
we have $\hat {p_1}\gamma_{\mu}=-\gamma_{\mu}\hat {p_1}+2p_{1\mu}$, so that:
\begin{equation}
\begin{array}{ll}
\displaystyle\frac{1}{4m}\overline{u}(p_2)\left [\gamma_{\mu}\hat{p_2}+\hat{p_1}\gamma_{\mu} \right ]u(p_1)&=\displaystyle\frac{1}{4m}\overline{u}(p_2)\left [-\hat{p_2}\gamma_{\mu}+2p_{2\mu}-\gamma_{\mu}\hat{p_1}+2p_{1\mu}\right ]u(p_1)\\
&=\displaystyle\frac{1}{4m}\overline{u}(p_2)\left [-2\gamma_{\mu}m+2(p_{2\mu}+p_{1\mu})\right ]u(p_1)\\
&=\displaystyle\frac{1}{2}\overline{u}(p_2)\left [-\gamma_{\mu}+\displaystyle\frac{(p_{2\mu}+p_{1\mu})}{m}
\right ]u(p_1).\\
\end{array}
\label{eq:apz}
\end{equation}
Inserting in Eq. (\ref{eq:apz}) in (\ref{eq:ap1}), we find Eq. (\ref{eq:apr}). 

Note in this respect, that the relation (\ref{eq:apr}) is correct only for the case when both nucleons are on mass shell, i.e; they are described by the four-component spinors $u(p)$, satisfying the Dirac equation. It is not the case for the quasi-elastic scattering of electrons by atomic nuclei, $a+A\to e+p+x$, which corresponds to the scattering $e+p^*\to e+p$, where $p^*$ is a virtual nucleon.

\section{Appendix II: relativistic description of the electron polarization}

Applying the Dirac equation to the four-component spinor $u(p)$, of an electron with mass $m_e$, one can find the following expression for the density matrix of polarized electrons:
\begin{equation}
\rho_{\alpha\beta}= u_{\alpha}(p) u^\dagger_{\beta}(p)
= \displaystyle\frac{1}{2}(\hat p+m_e)(1-\gamma_5\hat s),
\label{eq:a21}
\end{equation}
where $\alpha,~ \beta$ are the spinor indexes.

Here $s_\alpha$ is the four vector of the electron spin, which satisfies the following two conditions:
\begin{equation}
s\cdot p=0,~~s^2=-1
\label{eq:a22}
\end{equation}
In terms of the three-vector $\vec s$ of the electron polarization in rest, i.e. with zero three-momentum, the components of the four-vector $s_\alpha$ can be written as:
\begin{equation}
s_\alpha= \left (\vs +
\displaystyle\frac{(\vs\cdot \vp)\vp}
{m_e(\epsilon+m_e)}, \displaystyle\frac{\vs\cdot \vp}{m_e}\right )
\label{eq:a23}
\end{equation}

The condition $s^2=-1$ corresponds to full electron polarization, so $s^2=-\vs^2=-1.$

Eqs. (\ref{eq:a21}) and (\ref{eq:a23}) are essentially simplified in case of relativistic electrons, $\epsilon\gg m_e$. In this case:
\begin{equation}
s_\alpha= \displaystyle\frac{\epsilon}{m}s_\ell(\vu ,1)
\label{eq:a24}
\end{equation}
where $\vu$ denotes the unit vector along $\vp$ and $s_\ell=\vs\cdot\vp/|\vp|\equiv\lambda$.

Taking into account that for relativistic electrons:
\begin{equation}
p_\alpha= \epsilon(\vu,1),
\label{eq:a24a}
\end{equation}
it is possible to re-write Eq. (\ref{eq:a24}) in the form:
\begin{equation}
s_\alpha= \displaystyle\frac{p_{1\alpha}}{m}\lambda.
\label{eq:a25}
\end{equation}
Substituting Eq. (\ref{eq:a25}) into Eq. (\ref{eq:a21}), one can find the following expression for the density matrix of a relativistic polarized electron:
\begin{equation}
\begin{array}{ll}
 \rho&= 
 \displaystyle\frac{1}{2}(\hat p+m_e)\left (1-\gamma_5\displaystyle\frac{\hat p}
{m_e}\lambda\right )=
\displaystyle\frac{1}{2}(\hat p+m_e)+
\displaystyle\frac{\lambda}{2}(\hat p+m_e)
\displaystyle\frac{\hat p}{m_e}\gamma_5 \\
&= \displaystyle\frac{1}{2}(\hat p+m_e)+\displaystyle\frac{\lambda}{2}
\left ( p^2+m_e\hat p\right )\displaystyle\frac{1}{m_e}\gamma_5 \\
&=\displaystyle\frac{1}{2}(\hat p+m_e)(1+\lambda\gamma_5)\equiv 
\displaystyle\frac{1}{2}\hat p(1+\lambda\gamma_5),
\end{array}
\label{eq:a26}
\end{equation}
where we used the following property of the $\gamma_5$-matrix: 
$\hat p \gamma_5+\gamma_5\hat p=0$, for any four-vector $p_\alpha$.

Using this expression for the density matrix $\rho$, let us calculate the leptonic tensor $L_{\mu\nu}(\lambda)$, corresponding to the scattering of longitudinally polarized electrons (neglecting the electron mass):
\begin{equation}
L_{\mu\nu}(\lambda)=\displaystyle\frac{1}{2}Tr \gamma_{\mu}
\hat k_1(1+\lambda\gamma_5)\gamma_{\nu}\hat k_2=
\displaystyle\frac{1}{2}Tr\gamma_{\nu}\hat k_1\gamma_{\nu}\hat k_2+
\displaystyle\frac{\lambda}{2}Tr\gamma_{\nu}\hat k_1\gamma_5\gamma_{\nu}\hat k_2
=L_{\mu\nu}^{(0)}+\lambda L_{\mu\nu}^{(1)}.
\label{eq:a27}
\end{equation}
The tensor $L_{\mu\nu}^{(0)}$ corresponds to the scattering of unpolarized electrons:
\begin{equation}
L_{\mu\nu}^{(0)}=2k_{1\mu} k_{2\nu}+k_{1\nu} k_{2\mu}-g_{\mu\nu}k_1\cdot k_2
\label{eq:ell0}
\end{equation}

The tensor $L_{\mu\nu}^{(1)}$, describing the dependence on the longitudinal electron polarization can be written in the following form:
\begin{equation}
L_{\mu\nu}^{(1)}=\displaystyle\frac{1}{2}Tr \gamma_{\mu}\hat k_1\gamma_{\nu}
\hat k_2\gamma_5=-\displaystyle\frac{1}{2}Tr \gamma_{\mu}\gamma_{\nu}\hat k_1\hat k_2\gamma_5=2i\epsilon_{\mu\nu\rho\sigma}k_{1\rho}k_{2\sigma}
\label{eq:ell1}
\end{equation}
We applied another property of $\gamma_5$, that is:
$$Tr \gamma_{\mu}\gamma_{\nu}\gamma_{\rho}\gamma_{\sigma}\gamma_5=-4i
\epsilon_{\mu\nu\rho\sigma}.$$
Taking into account the conservation of four-momentum in the electron vertex: $k_1=k_2+q$, we can rewrite the tensor $L_{\mu\nu}^{(1)}$ in the following form, which is more convenient in this frame:
\begin{equation}
L_{\mu\nu}^{(1)}=2i\epsilon_{\mu\nu\rho\sigma}q_{\rho}k_{1\sigma}.
\label{eq:ell8}
\end{equation}
The three-vector $\vq $ has only nonzero $z-$component, in the Breit system. The tensor $\epsilon_{\mu\nu\rho\sigma}$ is defined in such way that $\epsilon_{xyz0}=+1$. 

Formulas (\ref{eq:ell1}) and (\ref{eq:ell8}) are very important to determine the sign of the polarization observables, in $ep$-elastic scattering.

\section{Appendix III: expression of  $\sin \displaystyle\frac{\theta_B}{2}$ in terms of kinematical variables in the Lab system.} 

Let us find the expression of $\sin \displaystyle\frac{\theta_B}{2}$ in terms
of kinematical variables in the Lab-system.

Using the relation (\ref{eq:cot}), one finds:
\begin{equation}
\displaystyle\frac{1}{\sin^2 \displaystyle\frac{\theta_B}{2}}=
1+\displaystyle\frac{\cot^2 \displaystyle\frac{\theta_e}{2}}{1+\tau}=
\displaystyle\frac{1}{1+\tau}\left [ \tau +\displaystyle\frac{1}{\sin^2 \displaystyle\frac{\theta_e}{2}}
\right ]=
\displaystyle\frac{1}{1+\tau}\displaystyle\frac{1+\tau \sin^2 \displaystyle\frac{\theta_e}{2}}{\sin^2 \displaystyle\frac{\theta_e}{2}}
\label{eq:a1}
\end{equation}
So
\begin{equation}
1+\tau\sin^2 \displaystyle\frac{\theta_e}{2}= 
1+\displaystyle\frac{\displaystyle\frac{\eu^2}{m^2}\sin^4 \displaystyle\frac{\theta_e}{2}}
{1+2\displaystyle\frac{\eu}{m}\sin^2 \displaystyle\frac{\theta_e}{2}}=
\displaystyle\frac
{(1+\displaystyle\frac{\eu}{m}\sin^2 \displaystyle\frac{\theta_e}{2})^2}
{1+2\displaystyle\frac{\eu}{m}\sin^2 \displaystyle\frac{\theta_e}{2}}.
\label{eq:a2}
\end{equation}

Using the relation (\ref{eq:e1e2}) between the initial and final electron energy, we have:
\begin{equation}
1+\displaystyle\frac{\eu}{m}\sin^2 \displaystyle\frac{\theta_e}{2}
=\displaystyle\frac{1}{2}\displaystyle\frac{\eu+\ed}{\ed}.
\label{eq:et}
\end{equation}
Substituting (\ref{eq:et}) in (\ref{eq:a1}), one finally finds:
\begin{equation}
\displaystyle\frac{1}{\sin^2 \displaystyle\frac{\theta_B}{2}}=
\displaystyle\frac{(\eu+\ed)^2}{(-q^2)(1+\tau)}.
\label{eq:ef}
\end{equation}

\end{document}